# Analyzing the Impact of Cognitive Load in Evaluating Gaze-based Typing

Korok Sengupta[1], Jun Sun[1], Raphael Menges[1], Chandan Kumar[1] and Steffen Staab[1,2]
[1]Institute for Web Science and Technologies (WeST)    [2]Web and Internet Science Research Group (WAIS)
University of Koblenz–Landau, Germany    University of Southampton, UK
Email: {koroksengupta, junsun, raphaelmenges, kumar, staab}@uni-koblenz.de

*Abstract*—Gaze-based virtual keyboards provide an effective interface for text entry by eye movements. The efficiency and usability of these keyboards have traditionally been evaluated with conventional text entry performance measures such as words per minute, keystrokes per character, backspace usage, etc. However, in comparison to the traditional text entry approaches, gaze-based typing involves natural eye movements that are highly correlated with human brain cognition. Employing eye gaze as an input could lead to excessive mental demand, and in this work we argue the need to include cognitive load as an eye typing evaluation measure. We evaluate three variations of gaze-based virtual keyboards, which implement variable designs in terms of word suggestion positioning. The conventional text entry metrics indicate no significant difference in the performance of the different keyboard designs. However, STFT (Short-time Fourier Transform) based analysis of EEG signals indicate variances in the mental workload of participants while interacting with these designs. Moreover, the EEG analysis provides insights into the user's cognition variation for different typing phases and intervals, which should be considered in order to improve eye typing usability.

*Keywords*-eye typing; gaze input; EEG; cognitive load

## I. INTRODUCTION

Eye movements provide a crucial pathway for interaction on digital platforms for people in locked-in state or impacted by neuromuscular problems, as they have very little or no control of muscles to operate a conventional mouse and keyboard combination. Eye gaze is feasible as an input source, as it is intuitive and requires only minor training [1] to adapt, especially since the "control-to-display" relationship is already established in the brain [2]. Therefore, gaze signals are used not only for pointing to emulate conventional mouse, but also for writing text and messages on virtual keyboards [3].

Eye typing is the process of entering text by gaze control, where the user selects letters on the virtual keyboard through looking at them. The process involves not only the primary task of scanning and selecting the letters to form words, but also reading through the collected inputs for validating the correctness. There has been different designs for gaze-based text entry that include input techniques like dwell time [4], and dwell free [5] based approaches. Another significant aspect is the exploitation of intelligent text prediction methods for more efficient text entry [6]. Moreover, the placement of word predictions [7] around the foveal region [8] has been investigated. However it is unclear, how these designs do impact the user cognition, i.e., if the mental effort required in text entry process varies for different designs.

Cognitive effort is often related to EEG signal analysis. EEG signals have been used in different experiments [9] – along with gaze signals – to navigate across different applications. Other directions are understanding artifacts caused by eye movements [10] or using EEG as event related potentials [11]. Yet, EEG have rarely been used to analyze gaze-based typing, although it might provide helpful feedback about the cognitive demand on the user.

In this paper, we analyze the EEG signals of users while they perform text entry via gaze. We investigate whether the analysis of EEG signals can help us to estimate the *cognitive load* associated with gaze-based typing, and provide information complementary to the conventional performance metrics. Our study indicates that the keyboard design can have significant impact on perceptive effort required from users, and hence measurement of cognitive load needs to be considered while evaluating eye typing interfaces. Cognitive load from EEG signals provides an insightful addition for interpretation of raw performance (e.g., words per minute, keystrokes saved, backspace usage). It can also be utilized to understand which part of text entry process leads to difficulty and requires attention to provide a better user experience.

## II. COGNITIVE LOAD

Antonenko *et al.* [12] define cognitive load as the load or the effort imposed on the memory by the cognitive processes involved in learning. This mental effort has been extended by Paas *et al.* [13] as the cognitive capacity that is allocated to take care of the demands imposed by the specific task under question. These research works – which focus on the cognitive architecture involving memory and time collectively – contribute towards a theory called *Cognitive Load Theory* (CLT) [13].

*Electroencephalogram* (EEG) signals provide a technique to investigate cognitive load, and are well studied over years [14]. Apart from EEG, *Galvanic Skin Response* (GSR) has also been used to estimate the effect of cognitive load [15]. Compared with other available options to measure the cognitive state of an individual such as PET, fMRI, fNRS; EEG has the advantage of both high temporal resolution [16] and economical flexibility. Development of low-cost and light-weight EEG devices like Emotiv EPOC[1] allows researchers to investigate the domain of cognitive load easily [17], thereby giving us further motivation to use such low-cost devices to study the cognitive reaction associated with gaze-based typing on virtual keyboards.

In this work, we apply *Short-time Fourier Transform* (STFT) to the EEG signal time series, in order to evaluate the cognitive load of each participant during the experiments [18]. Compared with simple *Fast Fourier Transform* (FFT), STFT is able to capture both time and frequency information in non-stationary signals – as in our case. We executed following pipeline to extract cognitive load out of the raw EEG channels' data:

*(1) Preprocessing.* We first divide each signal time series into multiple sliding windows of equal length ($1024$ samples, or $8$ seconds) with a window slide unit of length $512$ samples ($4$ seconds). This results in two neighboring windows sharing an overlap of $50\%$ window length. A half-cosine window function is then applied to each windowed signal in order to minimize the effect of leakage [19].

*(2) Fourier transform.* In this step, a discrete Fourier transform of each windowed signal $c$ with length $N=1024$ and sampling rate $Fs=128$Hz is performed (see Equation 1) [20], resulting in its spectrogram. This is subdivided into frequency bands known as *Delta* ($<4$Hz), *Theta* ($\geq 4$Hz and $<8$Hz), *Alpha* ($\geq 8$Hz and $<14$Hz) and *Beta* ($\geq 14$Hz), according to frequency [21].

$$C_k = \sum_{j=0}^{N-1} c_j e^{2\pi ijk/N} \qquad k = 0, ..., N-1 \qquad (1)$$

*(3) Computation of spectral power.* For a defined frequency band $[f_1, f_2]$, we can further estimate its *spectral power* $P$ (see Equation 2) and the corresponding *spectral power ratio* (spectral power in a certain frequency band divided by total power in all bands).

$$P = \frac{1}{N} \sum_k |C_k|^2 \qquad k \in [\lfloor f_1 \cdot N/Fs \rfloor, \lfloor f_2 \cdot N/Fs \rfloor] \qquad (2)$$

Studies have shown that for experiment participants who are performing certain tasks with higher cognitive load (e.g., writing) – compared with relaxing – a higher percentage of high frequency EEG waves (especially in the Beta band) can be observed [22]. Hence, we can compute the average value of the spectral power ratio of the Beta band in the EEG signal from all 14 channels within a time window. This servers as an indicator of the cognitive load *within this particular time window*.

### III. A COMPARATIVE STUDY OF GAZE-BASED KEYBOARDS

The success and usability of word predictions highly depends on the presentation and user interface parameters [23],

[1] https://www.emotiv.com/epoc 128Hz Sampling Rate, 14 Channels

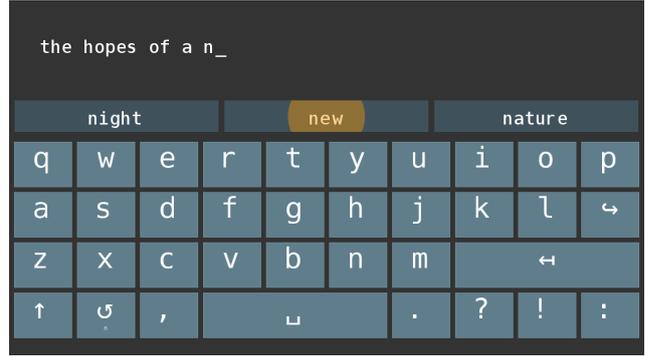
(a) Keyboard A with top line of suggestion

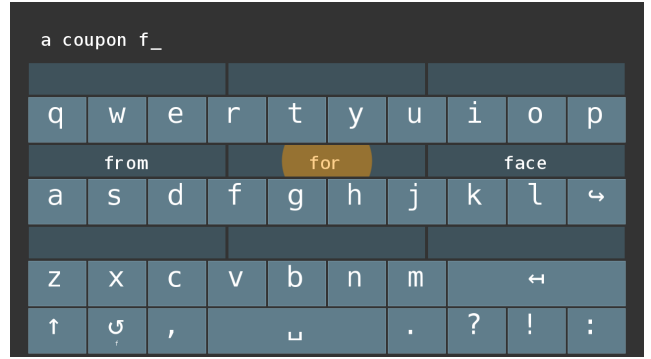
(b) Keyboard B with inter-spaced line of suggestions

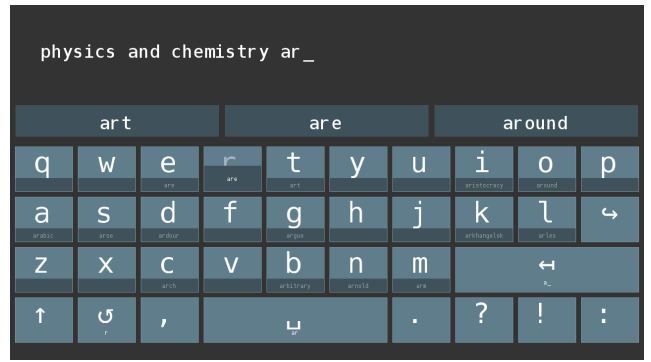
(c) Keyboard C with bot top line of suggestions and per key

Figure 1. Three keyboard designs of our experiment.

which includes the number of suggestions to display and of course the positioning of suggestions in the screen space of the keyboard. The positioning is a crucial aspect since it deals with visual attention of the user while typing letters and relates to the cognitive and perceptual influence. In touch-based virtual keyboards, several approaches using variable positioning like inter-spaced and on-key predictions to reduce the effect of visual scan and search time have been evaluated [24]. The concept of variable positioning have also been vaguely analyzed for eye-based interactions [25]. In the most conventional gaze-based keyboard designs, the list of predicted words is placed on top of the key layout near the text entry area [6]. In the *AugKey* approach [8], word suggestions are placed at the right side of the keyboard and it also includes pre- and postfixes around the focused key to exploit the foveal region of visual perception.

The above-mentioned approaches do emphasize the role of word prediction in gaze-based text entry. However, until now it is not evident whether the variable positioning of these predictions have any impact by means of reducing eye-movements, visual search or scanning time. Researchers did present some arguments of word predictions being related to the additional cost of perceptual and cognitive load, caused by shifting the focus from the keyboard to the word list and the repeated scanning of the list [26]. There have been no concrete studies to investigate if the variable positioning of word predictions has a correlation with the visual attention and can enhance the user experience while typing. In the eye typing evaluation studies, one of the most important and widely used metrics are *words per minute*. This metric has been consistently the benchmark to measure the performance of keyboards and how design choices and technology influence on the typing speed. Thus, to understand the role of cognitive load in eye typing, we devised an experiment to observe the relation of the cognitive load to the impact of suggestions across different positions.

The conventional design that is prevalent across most virtual keyboard is represented in *Keyboard A* (Figure 1a). It has a single line of word predictions on top of the keys layout. *Keyboard B* (Figure 1b) integrates the suggestions inter-spaced within the key rows. This design is the first step towards integrating suggestions with the keyboard layout. *Keyboard C* (Figure 1c) [27] includes additional suggestions on the keys apart from the conventional top level suggestion row, like in Keyboard A.

## IV. METHODOLOGY

Each participant was asked to be a part of three sessions on three different days, each dedicated to one keyboard design. The experiment was executed in a controlled environment with artificial illumination. Latin square ordering was used for the counter-balanced setup of experimental session slots. The the independent and control variables were carefully noted prior to the experimental process. Each participant was instructed on how the experimental process will be carried out in a short training session. They were specifically trained to read the sentence to type. The participants were also instructed on how the hinted sentence disappears on the selection of the first letter. This behavior was chosen to ensure the simulation of free writing and to prohibit the participant from comparing the collected input letter-per-letter, which would influence eye gaze data [28].

### A. Participants

The main experiment had five able bodied male participants who were paid to participate; age ranging from 22 to 26 years (mean = 24.2, SD = 2.17). None of the participants had prior eye typing experience and none of them wore any sort of corrective visual devices. Every participant was conversant in the QWERTY layout, which was used in the designs.

### B. Apparatus

The SMI REDn eye tracker was set up with sampling rate of 60Hz to collect the gaze data. The eye tracker was attached to a 24 inch monitor. The participant was placed in a height-adjustable chair. For the BCI device, Emotiv's 14 channel EPOC+ device was chosen to measure the brain signals at a sampling rate of 128Hz. The Premium SDK allowed us to extract the raw EEG data of each channel.

### C. Implementation

The three different keyboard designs were implemented as eyeGUI [29] elements. The eyeGUI library is programmed in C++ and offers various user interface elements for gaze-controlled applications. Most basic features like an OpenGL abstraction, Unicode encoding and font rendering are included and have been used for the keyboards. For both text and letters on the keys, the DejaVUSans Mono[2] has been chosen, as it offers a clean typeface and is freely available. Bitmaps from the font are generated with the FreeType2[3] library and rendered onto a texture atlas. For every letter within a word, a rectangular geometry is generated and the appropriate part of the texture atlas containing the desired letter is rendered on this quad with a shader program. The displayed suggestions are generated by the Presage[4] prediction engine. We took 50,000 random well formed English sentences from the Tatoeba.com[5] database for the dictionary learning. The words from the sentences to be typed by the participant were added in random order. If not added in random order, the prediction engine would suggest the sentences of the task after a few typed characters.
Recording and synchronization of keyboard event markers, eye tracking and brain data was achieved with LabStreamingLayer[6], which provided us with synchronized time stamps.

### D. Procedure

For each participant, there were five sessions for each keyboard, and each session had five sentences, plus a single training session in the beginning of the experiment. The sentences were taken from the phrase set of Mackenzie and Soukoreff [30]. Each sentence to type in was presented in the text display at the top of the keyboard layout and vanished after the first key stroke. Then, the collected input was presented in this area.

---
[2] https://dejavu-fonts.github.io
[3] http://www.freetype.org
[4] http://presage.sourceforge.net
[5] https://tatoeba.org/eng
[6] https://github.com/sccn/labstreaminglayer

Each participant was requested to sign the informed consent form prior to their experimental session. They were then given a proper description of the experiment and the devices being used. Prior to every session, the eye tracker was calibrated. In the training session, they were shown that in order to submit a typed sentence, they needed to hit the space-bar on the computer's physical keyboard.

## V. RESULTS AND OBSERVATIONS

The experimental results provide an indication on the significant role of mental workload assessment while performing the high cognitive agility task of eye typing. In the following subsections, we first provide the details on conventional performance metrics, then we present the experimental results on cognitive load and discuss our findings.

### A. Performance

The metric *words per minute* (wpm) [31] is one of the most basic and widely used metric for performance evaluation of eye typing experiments. For this analysis it has been calculated as: $((|T|-1)*60) \div (5*s)$ [31], where $|T|$ is the length of the transcribed string and $s$ is the time taken to transcribe the text in seconds, including backspaces. The average number of letters per word is approximated with 5, to normalize the input sentences. Based on wpm, the grand means of each of the keyboards are very close to one another: 9.20, 8.60 and 9.05 wpm for Keyboard A, B and C, respectively. ANOVA for wpm values reveal a non-significant effect with $F(2,12)=0.403$, $p>0.05$.

*Keystrokes per character* (KPSC) [32] is another standard metric which is often used. We have adapted this concept to measure how much *keystrokes* were *saved* during a session. This reveals how word suggestions are influential in reducing the typing effort, thereby reducing the time required for typing. The average percentage of keystrokes saved was 39.0018, 35.4366, 33.4694 for the three keyboard setups. ANOVA for keystrokes saved however indicates a non-significant effect, with $F(2,12)=1.54$, $p>0.05$.

The *backspace key usage* is another indicative metric that hints about the amount of mistakes rectified by the users while typing. Since eye typing is an exhaustive task, people often make mistakes. The average usage of backspace for the three keyboards A, B and C were 2.92, 6.32 and 5.00 times. ANOVA for backspace key usage indicate a non-significant effect, with $F(2,12)=1.64$, $p>0.05$.

### B. Cognitive Load

In this section, we compare the cognitive load of our participants in different experiment setups. As discussed in Section II, we use the spectral power ratio of Beta band of EEG signals to indicate the level of cognitive load.

Figure 2 shows the average cognitive effort required by participants for different keyboards. We can observe that

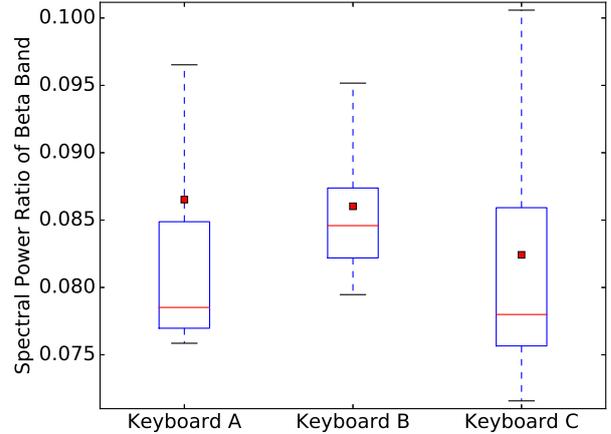

Figure 2. Comparison of overall cognitive load of participants using the three keyboards during the experiment. The X-axis marks the keyboard; while the Y-axis denotes the spectral power ratio of the Beta band of EEG signals, which indicates the level of cognitive load of the participant. Each data entry in the box-plot corresponds to the spectral power ratio value of one *time window* (see Section II). The horizontal bar in the middle of the box shows the median value, while the red dot shows the mean value (same for all boxplots in this paper).

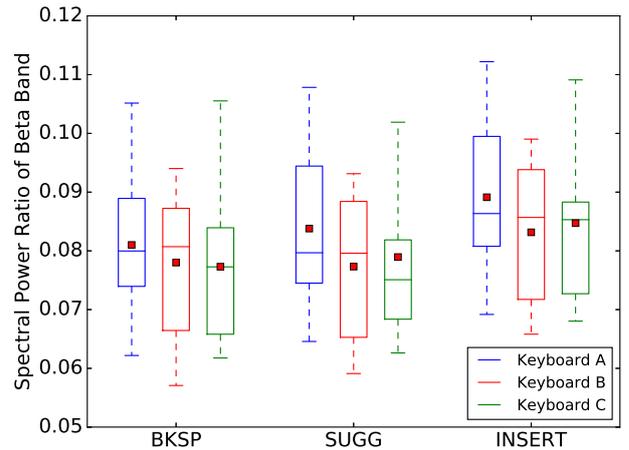

Figure 3. Comparison of cognitive load of participants in different typing modes using the three keyboards (shown with different colors) during the experiment. The X-axis labels different modes; while the Y-axis shows the spectral power ratio of the Beta band of EEG signals, which indicates the level of cognitive load of the participant. In each boxplot we have 25 samples in total (outliers are omitted), corresponding to the 5 participants and 5 sessions for each participant.

Keyboard C (with mean value 0.0824) has lesser cognitive load compared to both Keyboard A (with mean value 0.0865) and B (with mean value 0.0860). T-test shows that the differences are significant (Keyboard A and C with p=0.01542, N=150; Keyboard B and C with p=0.00047, N=150)[7]. Keyboard C embeds individual suggestions on the letters itself and hence users might have required less cognitive effort of scanning word list. Keyboard B has similar predictions as A, however the dynamic appearance of inter-spaced word list seems to confuse users, leading to high mental demand. Some participants in a parallel experiment also revealed similar observations, as they stated design B as frustrating, however design C to be more consistent.
These results do not have a direct correlation with the conventional performance metrics, however all three metrics in Section V-A indicate a lower performance for Keyboard B (non-significant), and EEG analysis indicating significantly higher cognitive load, implying Keyboard B as a bad design choice for end users.

---
[7]N=5 participants * (1 training + 5 experimental sessions) * 5 sentences

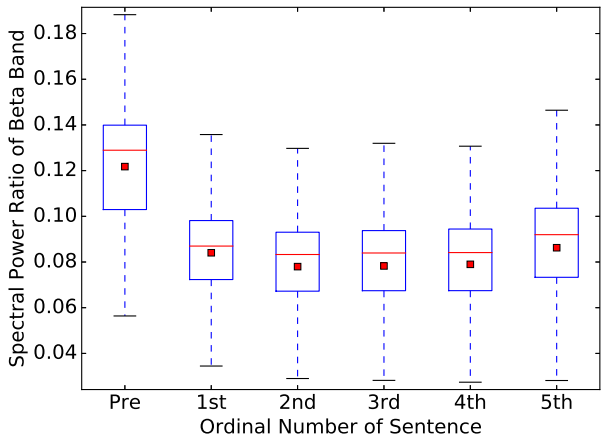

Figure 4. Comparison of cognitive load of participants when typing different sentences using Keyboards C during the experiment (the other two keyboards provide a similar pattern). The X-axis contains the ordinal number of the sentence in each session (`Pre` is the time before a keyboard is displayed and a participant asked to type); while the Y-axis shows the spectral power ratio of the Beta band of EEG signals, which indicates the level of cognitive load of the participant. In each boxplot we have 25 samples in total (outliers are omitted), corresponding to the 5 participants and 5 sessions for each participant.

We were also keen to investigate how different aspects of text entry process impact user cognition. Hence we compared the cognitive load of participants in different typing *modes* when using the three keyboards:

- `BKSP`: the participant is deleting content by hitting the backspace key on the eye tracking keyboard
- `SUGG`: the participant is selecting the suggestions provided by the eye tracking keyboard
- `INSERT`: the participant is inserting single letters

Figure 3 reveals that the cognitive load is lower for all designs when the participants were deleting content (`BKSP`) or using suggestions on the keyboard (`SUGG`), than inserting content letter by letter (`INSERT`). This indicates a higher demand while selecting letters, as one needs to scan and process the information in the foveal region and then finalize which one to pick. However, when deleting letters because of a mistake, one needs to repeatedly fixate the backspace key. It could also relate to why Keyboard A, despite having less backspace usage and errors, does not perform better. The effort required for error correction has no major impact.

Furthermore, we compared the cognitive load when the participants were in the pre-experiment phase (the time between each section starts and the first sentence is shown) to the task of typing the sentences. As shown in Figure 4, we do not observe significant difference of the cognitive load among different sentences within a session. This could be explained by the fact that we randomized the order of different sentences, as some sentences might be more cognitive demanding than others. However, we do observe that the cognitive load during the pre-experiment phase is higher than the one during actual typing. This indicates that users need time to adjust themselves to the experimental eye typing environment, but once they got used to the environment, the cognitive load is rather stable.

## VI. CONCLUSION

In this paper, we conducted a small scale experimental study to analyze the impact of cognitive load on gaze-based typing. We performed an assessment of virtual keyboards with variable positioning of word predictions. The results indicate the need to assess cognitive load impact in gaze-based typing scenarios. It provides us a valuable direction to understand gaze-based keyboard designs from the perspective of cognitive load. The alternation in word prediction positioning creates little or no effect on traditional performance metrics but according to the EEG analysis, it is quite evident that cognitive load varies. In future, we aim to improve the eye typing usability by adapting the dwell time of virtual keyboards based on instantaneous cognitive load.

## ACKNOWLEDGMENT

This work is part of project MAMEM that has received funding from the European Union's Horizon 2020 research and innovation program under grant agreement number: 644780.